\documentclass[aps,twocolumn,preprintnumbers,amsmath,amssymb,nofootinbib,superscriptaddress,notitlepage]{revtex4-1}

\usepackage{epsfig}
\usepackage{slashed}

\usepackage[utf8]{inputenc}

\begin{document}

\title{Exotic Doubly Charmed $D_{s0}^*(2317)\,D$ and $D_{s1}^*(2460)\,D^*$ Molecules}

  \author{M. Sanchez Sanchez}
  \affiliation{Institut de Physique Nucl\'eaire, CNRS-IN2P3, Univ. Paris-Sud, Universit\'e Paris-Saclay, F-91406 Orsay Cedex, France}
  \author{Li-Sheng Geng}\email{lisheng.geng@buaa.edu.cn}
  \author{Jun-Xu Lu}
  \affiliation{School of Physics and Nuclear Energy Engineering, \\
International Research Center for Nuclei and Particles in the Cosmos and \\
Beijing Key Laboratory of Advanced Nuclear Materials and Physics, \\
Beihang University, Beijing 100191, China}
  \author{Tetsuo Hyodo}
  \affiliation{Yukawa Institute for Theoretical Physics, Kyoto University,
606-8502, Japan}

  \author{M. Pavon Valderrama}\email{mpavon@buaa.edu.cn}
  \affiliation{School of Physics and Nuclear Energy Engineering, \\
International Research Center for Nuclei and Particles in the Cosmos and \\
Beijing Key Laboratory of Advanced Nuclear Materials and Physics, \\
Beihang University, Beijing 100191, China}

\date{\today}


\begin{abstract} 
\rule{0ex}{3ex}
The $D_{s0}^*(2317) D$ and $D_{s1}^*(2460) D^*$ heavy meson systems
can exchange a kaon that is emitted in S-wave owing to the opposite
intrinsic parity of the $D_{s0}^*$($D_{s1}^*$) and $D$($D^*$) mesons.
As a consequence of the mass difference of the $D_{s0}^*$($D_{s1}^*$)
and $D$($D^*$) mesons, the range of the kaon exchange potential
will be significantly longer than expected, corresponding
to an effective mass of about $200\,{\rm MeV}$.
The potential will be very strong: the strength of the interaction
is proportional to $(m_{D_{s0}} - m_D)^2 / f_{\pi}^2$ and
$(m_{D_{s1}} - m_{D^*})^2 / f_{\pi}^2$.
This combination of range and strength almost guarantees the existence of
$D_{s0}^*(2317) D$ and $D_{s1}^*(2460) D^*$ bound states with $J^P = 0^{-}$
and $J^P = 0^{-}, 2^{-}$ respectively.
Concrete calculations indicate a binding energy of $5-15\,{\rm MeV}$
independently of $J^P$.
The $D_{s0}^*(2317) D$ and $D_{s1}^*(2460) D^*$ molecules have manifestly
exotic flavour quantum numbers: $C=2$, $S=1$ and $I=1/2$.
We expect the existence of bottom counterparts composed of
the $B B_{s0}$ and $B^* B_{s1}^*$ mesons, which will be more bound and
have a richer spectrum that might include a shallow P-wave state
and an excited S-wave state.
\end{abstract}

\maketitle

The existence of hadronic molecules was conjectured
long ago~\cite{Voloshin:1976ap,DeRujula:1976qd}
on the basis of a very simple idea: the exchange of light mesons 
between two hadrons generates a potential that might be able
to bind them.
The discovery of the $X(3872)$ by Belle~\cite{Choi:2003ue}
more than a decade ago provided a very strong candidate for a molecular state,
owing to its small width and to its closeness
to the $D^0 \bar{D}^{0*}$ threshold.
Subsequently other molecular candidates have been discovered, among them
the $Z_c$'s~\cite{Ablikim:2013mio,Liu:2013dau}
(which are conjectured to be $D\bar{D}^*$ and $D^*\bar{D}^*$
molecules~\cite{Wang:2013cya,Guo:2013sya}),
the $Z_b$'s~\cite{Belle:2011aa,Adachi:2012im}
($B\bar{B}^*$ and $B^*\bar{B}^*$ molecules~\cite{Voloshin:2011qa,Cleven:2011gp})
and the $P_c(4450)$ pentaquark-like state~\cite{Aaij:2015tga}
(a $\Sigma_c^* \bar{D}^*$~\cite{Chen:2015loa}
or a $\Sigma_c \bar{D}^*$ molecule~\cite{Karliner:2015ina,Chen:2015moa,Roca:2015dva,Xiao:2015fia},
in the latter case probably with a sizable $\Lambda_c(2590) \bar{D}$
component~\cite{Burns:2015dwa,Geng:2017hxc}).

Though it is easy to conjecture the existence of hadronic molecules 
from theoretical principles, making concrete predictions
is considerably harder.
The reason is that in most cases hadronic molecules are generated
as a consequence of unknown short-range physics.
This is manifest from the necessity of cut-offs / form-factors.
If we consider the one pion exchange (OPE) potential, which is expected
to be the longest range piece of the interaction between two hadrons
(provided they contain at least one light quark),
we will quickly realize that it requires regularization:
the OPE potential contains a tensor piece that is singular at short distances.
The tensor force, if attractive, will be able to hold an infinite number of
bound states.
This situation is circumvented by the introduction of a form factor,
cut-off or other regulator that renders predictions possible
at the price of the introduction of an unknown new parameter~\cite{Tornqvist:1993ng,Ericson:1993wy,Liu:2008fh,Thomas:2008ja,Sun:2012sy,Li:2012ss}.
Educated guesses are possible by making a judicious choice of
the cut-off, the work by T\"ornqvist on heavy meson-antimeson
bound states being an astonishing example~\cite{Tornqvist:1993ng},
but there always remains a large degree of arbitrariness.

Yet the tensor force is not present in every hadron molecule.
The richness of the hadron spectrum gives rise to other possibilities
even if we only consider the exchange of a pseudo Nambu-Goldstone boson.
If a pion or a kaon is exchanged in a vertex involving hadrons
with different parities a series of interesting situations can arise.
If in addition a vertex involves hadrons with different masses,
this can lead to interactions with an unusual long range
for strong interactions.
A recent example is a Coulomb-like force in the $\Lambda_c(2590){\Sigma}_c$ and
$\Lambda_c(2590)\bar{\Sigma}_c$ systems~\cite{Geng:2017jzr}.
In this case we have a $1/r$ potential that is regular at short distances:
we can make predictions that do not crucially depend on a cut-off,
though we still expect the unknown details of the short-range physics
to have a certain impact.

Here we consider the $D D_{s0}^*(2317)$ and $D^* D_{s1}^*(2460)$ systems,
which share interesting similarities with the $\Lambda_{c}(2590) \Sigma_c$
molecule theorized in Ref.~\cite{Geng:2017jzr}.
The $D$ and $D^*$ mesons have negative parity, while for the $D_{s0}^*(2317)$
and $D_{s1}^*(2460)$ mesons ($D_{s0}^*$ and $D_{s1}^*$ from now on)
the parity is positive.
That is, they can exchange a kaon in S-wave.
In addition the mass difference $m(D_{s0}^*) - m(D)$ and $m(D_{s1}^*) - m(D^*)$ 
is similar to the kaon mass, which means that the exchanged
kaon will be near the mass shell and hence the range of
the interaction will be unusually large.
If this were not enough, chiral symmetry implies that the strength of
the $D_{s0}^* D K$ and $D_{s1}^* D^* K$ vertices are proportional to
the mass difference, which translates into an exceptional strength
for the resulting Yukawa potential.
This mechanism is also present in the $\Lambda(1405) N$ and
$\Xi(1690) \Sigma$ via antikaon exchange and
in the $\Lambda(1520) \Sigma^*$ via pion exchange.

This type of kaon exchange leads to a different spectrum than the one
obtained from standard OPE~\cite{Tornqvist:1993ng}.
The strength of the former is independent of spin and isospin,
while the later is proportional to
$\pm \, \vec{S}_1 \cdot {S}_2 \,\vec{T}_1 \cdot {T}_2$,
with $\vec{S}_{1,2}$ and $\vec{T}_{1,2}$ the spin and isospin of
hadrons $1$ and $2$ and the $+/-$ sign
for hadron-hadron/hadron-antihadron~\cite{Karliner:2015ina}.
For standard pion exchange flavour exotic states are suppressed as they
require a symmetric/antisymmetric wave function. This limits
the choices of total spin and isospin for which attraction is strong.
In turn CP exotic states are suppressed as they usually require P-wave,
for which binding is less likely.
The type of kaon exchange discussed here is independent of spin
(the kaon is emitted in S-wave) and of isospin
(the $D_{s0}^*$ and $D_{s1}^*$ are isoscalars).
We nonetheless stress that the production of flavour exotic states
is experimentally difficult and has only been achieved
recently~\cite{Aaij:2017ueg}.

The $D D_{s0}^*(2317)$ and $D^* D_{s1}^*(2460)$ molecules are also interesting
for another reason: their quark content is $cc {\bar s} {\bar q}$ with
$\bar q = \bar u$, $\bar d$.
This configuration is unlikely to form compact tetraquarks
but narrow molecules instead,
as argued by Manohar and Wise~\cite{Manohar:1992nd}.
Lattice QCD~\cite{Bicudo:2015vta,Francis:2016hui} and
quark model calculations~\cite{Carlson:1987hh,Gelman:2002wf,Vijande:2009kj,Du:2012wp,Luo:2017eub,Karliner:2017qjm,Czarnecki:2017vco}
seem to indicate that compact $QQ \bar q \bar q$ structures
exist in the bottom sector, but not in the charm one
(maybe with the exception of an isoscalar $cc \bar u \bar d$
state with $J^P = 1^+$).
As a consequence the potential discovery of a structure
with $cc {\bar s} {\bar q}$ quark-content and negative parity
will unmistakably point to a molecule. 

Now we calculate the one kaon exchange (OKE) potential
in the $D D_{s0}^*$ and $D^* D_{s1}^*$ molecules.
We begin with the ${D, D^*}$ S-wave heavy mesons,
which can be written as the heavy quark symmetric superfield: 
\begin{eqnarray}
H_a = \frac{1+\slashed v}{2}\,
\left[ D_a^{*\mu} \gamma_{\mu} - D_a \gamma_5 \right] \, ,
\end{eqnarray}
where $a$ is an SU(3)-flavour index such that
\begin{eqnarray}
D_a =
\begin{pmatrix}
D^{0} \\
D^{+} \\
D_s
\end{pmatrix} \, , \, 
D_a^* =
\begin{pmatrix}
D^{*0} \\
D^{*+} \\
D_s^*
\end{pmatrix} \, .
\end{eqnarray}
If we consider now the $D_0$ and $D_1$ P-wave heavy mesons (to which the 
$D_{s0}^*$, $D_{s1}^*$ belong), they can be arranged in the superfield
\begin{eqnarray}
S^a = \frac{1+\slashed v}{2}\,
\left[ D_{1}^{a\mu} \gamma_{\mu} \gamma_5 - D^a_{0} \right] \, ,
\end{eqnarray}
with the SU(3)-flavour structure
\begin{eqnarray}
D_{0}^a =
\begin{pmatrix}
D_0^{0} \\
D_0^{+} \\
D_{s0}^*
\end{pmatrix} \, , \, 
D_{1}^{a} =
\begin{pmatrix}
D_1^{0} \\
D_1^{+} \\
D_{s1}^*
\end{pmatrix} \, .
\end{eqnarray}
While the $D_{s0}^*$ and $D_{s1}^*$ are narrow and thus good candidates
for being part of a molecule,
the $D_0^{0}$, $D_0^{+}$, $D_1^{0}$ and $D_1^{+}$ are broad
($\Gamma \sim 200-300\,{\rm MeV}$) and as a consequence
unlikely to form bound state, except with kaons~\cite{Guo:2011dd}.
Besides the $D_{s0}^*$ and $D_{s1}^*$ are expected to contain a non-negligible
$D K$ and $D^* K$ molecular component~\cite{Guo:2006fu,Guo:2006rp,Guo:2011dd}
(about $50-70\%$ according to Refs.~\cite{Torres:2014vna,Ortega:2016mms})
plus a $D_s \eta$ and $D_s^* \eta$ component~\cite{Altenbuchinger:2013vwa}.
The binding momentum is about $200\,{\rm MeV}$ for the $D K$ and $D^* K$
and about $400\,{\rm MeV}$ for the $D_s \eta$ and $D_s^* \eta$.
If the binding momentum  of a $D D_{s0}^*$ / $D^* D_{s1}^*$ molecule is
smaller than these figures, it will be safe to ignore the possible
compound structure of the $D_{s0}^*$ and $D_{s1}^*$ mesons.

The heavy meson chiral lagrangian for the interaction
between the S- and P-wave heavy mesons is~\cite{Falk:1992cx}
\begin{eqnarray}
\mathcal{L} = \frac{h}{2}
\,{\rm Tr}\left[ \bar{H}_a S_b {\slashed A}_{ab} \gamma_5
\right] + {\rm H.C.}\, ,
\end{eqnarray}
with $a$, $b$ SU(3)-indices, $A^{\mu}_{ab}$ the axial current of the
pseudo Nambu-Goldstone field and where ${\rm H.C.}$
indicates the hermitian conjugate.
We have $A^{\mu} = - \frac{1}{f_{\pi}}\partial_{\mu} M$
with $f_{\pi} \simeq 130\,{\rm MeV}$, where
\begin{eqnarray}
M = 
\begin{pmatrix}
\frac{\pi^0}{\sqrt{2}} + \frac{\eta}{\sqrt{6}} & \pi^{+} & K^{+} \\
\pi^{-} & - \frac{\pi^0}{\sqrt{2}} + \frac{\eta}{\sqrt{6}} &  K^{0} \\
K^{-} & \bar{K}^0 & - \sqrt{\frac{2}{3}}\,\eta
\end{pmatrix} \, .
\end{eqnarray}
We determine the coupling $h$ from two different assumptions of
the structure of $D_{s0}^*$ and $D_{s1}^*$ states.
In the first scenario, assuming that the $D_{s0}^*$ and $D_{s1}^*$ are $c\bar{s}$
states, we can infer $h$ from the decays of $D_0$ and $D_1$ mesons
\begin{eqnarray}
\Gamma(D_0 \to D \pi) &=& \Gamma(D_0 \to D \pi^0) + \Gamma(D_0 \to D \pi^{\pm}) 
\nonumber \\
&=& \frac{3}{2}\,\Gamma(D_0 \to D \pi^{\pm}) \nonumber \\
&=& \frac{3}{2}\,\frac{m_D}{m_{D_0}}\,
\frac{q_{\pi}}{2 \pi}\,\frac{h^2}{f_{\pi}^2}\,(m_{D_0} - m_D)^2
\, ,
\end{eqnarray}
plus the analogous formula for the $D_1 \to D^* \pi$ decay, where
$q_{\pi} = \sqrt{(m_{D_0} - m_D)^2 - m_{\pi}^2}$ is the momentum of the pion.
If the widths of the $D_0$ and $D_1$ heavy mesons
are saturated by the pion decays above, we obtain $h \sim 0.5-0.9$
where the large spread comes from the uncertainties in the masses and
widths of the $D_0$ and $D_1$ mesons and also because it depends on
whether we use the $D_0^0$, the $D_0^{+}$ or the $D_1^0$ decay width
(notice that the $D_1^{+}$ has not been detected yet).
For instance Ref.~\cite{Colangelo:2012xi} obtains the values
$h = 0.61 \pm 0.07$, $0.50 \pm 0.06$ and $0.8 \pm 0.2$
for the three previous cases.
Determinations of this coupling
from QCD sum rules~\cite{Colangelo:1995ph,Colangelo:1997rp}
and lattice QCD~\cite{Becirevic:2012zza}
lie in the previous range.
That is, if the $D_{s0}^*$ and $D_{s1}^*$ are compact $c \bar{s}$ states
the uncertainty in $h$ is likely to be large,
for instance $h = 0.7 \pm 0.2$.

In the second scenario we deduce $h$ from the molecular hypothesis,
where the $D_{s0}^* \to D K$ ($D_{s1}^* \to D^* K$) coupling $g$
is extracted from the residues of the scattering amplitude
at the pole~\cite{Gamermann:2007fi,Torres:2014vna}.
We have the relation
\begin{eqnarray}
  g = \sqrt{2 m_{D_{s0}^*} 2 m_D} \, \omega_K \frac{h}{f_{\pi}} \, ,
\end{eqnarray}
from which a typical $g \sim 10-12\,{\rm GeV}$~\cite{Gamermann:2007fi,Torres:2014vna} translates into $h \sim 0.7-0.8$, where the higher value comes
from analyzing lattice QCD data~\cite{Lang:2014yfa}.
In this scenario we can use $h = 0.7 \pm 0.1$ for $f_{\pi} = 130\,{\rm MeV}$.
Choosing $f_{K} = 160\,{\rm MeV}$ instead of $f_{\pi}$ only amounts
to the change $h = 0.9 \pm 0.1$. This makes no difference
because the potential is proportional to $g^2$. 

The leading order (${\rm LO}$) potential for the $D D_{s0}^*$
and $D^* D_{s1}^*$ system is generated from kaon exchange
and is not diagonal.
If we consider the bases $\{ D D_{s0}^*, D_{s0}^* D \}$
and $\{ D^* D_{s1}^*, D_{s1}^* D^* \}$,
the momentum space potential reads
\begin{eqnarray}
V_{\rm OKE}(\vec{q}) = -h^2 \frac{\omega_K^2}{f_{\pi}^2}
\frac{1}{m_K^2 - \omega_K^2 + \vec{q}^2}
\begin{pmatrix}
0 & 1 \\
1 & 0
\end{pmatrix} \, ,
\end{eqnarray}
with $\omega_K = m_{D_{s0}^*} - m_D$ or $m_{D_{s1}^*} - m_{D^*}$
depending on the case.
We have used $f_{\pi}$ instead of $f_{K}$ as they are only different
at next-to-leading order in the chiral expansion.
The interesting point is that the range of the potential
is set by the effective kaon mass $\mu_K$
\begin{eqnarray}
\mu_K^2 = m_K^2 - \omega_K^2 \, , 
\end{eqnarray}
which is about $200\,{\rm MeV}$, moderately long-ranged.
This enhanced range also happens
in $\Lambda(1405) N$~\cite{Uchino:2011jt}.
For $S=0,2$ we have the linear combinations
$[| D D_{s0}^* \rangle + | D_{s0}^* D \rangle ] / \sqrt{2}$ and
$[| D^* D_{s1}^* \rangle + | D_{s1}^* D^* \rangle ] / \sqrt{2}$
for which the potential is attractive~\footnote{This is a consequence
  of extended Bose-Einstein statistics. The potential exchanges the $D$ by
  the $D_{s0}^*$ and vice versa, which means that it is convenient to
  consider the $D$ and $D_{s0}$ as identical particles. Alternatively
  we can notice that the potential is defined in the $D D_{s0} \to D D_{s0}$
  channel, which leads to an overall $(-1)^S$ factor (see for instance
  Ref.~\cite{Uchino:2011jt}).
}
and reads as
\begin{eqnarray}
V_{\rm OKE}(r) = - h^2 \frac{\omega_K^2}{f_{\pi}^2}\,\frac{e^{-\mu_K r}}{4 \pi r} \, ,
\end{eqnarray}
in configuration space, which has bound states if 
\begin{eqnarray}
\lambda_B = \frac{2 \mu_H}{\mu_K}\,
\frac{\omega_K^2}{4 \pi f_{\pi}^2}\,h^2 \geq 1.68 
\end{eqnarray}
with $\mu_H$ the reduced mass of the system.
This condition is probably satisfied:
the evaluation of the expression above yields $9.16 \, h^2$ and $10.70 \, h^2$
for the $D D_{s0}^*$ and $D D_{s1}^*$ cases respectively and
a bound state exists for $| h | > 0.43$ and $0.40$.
For $\lambda_B \geq 6.45$ there will be two bound states,
a condition that requires $| h | > 0.84$ and $0.78$,
which makes the existence of the second state less probable
but still possible.

Concrete calculations of the binding will be divided in two scenarios:
a compact and a molecular $D_{s0}^*$/$D_{s1}^*$.
In the first case the predictions will be subjected to large errors owing
to the poor knowledge of the coupling $h$.
In the second the coupling $h$ is well determined,
but the finite size of the $D K$ and $D^* K$ molecule has to be considered.
The OKE potential is regular but its short-range behaviour is not
necessarily physical.
We regularize it in order to obtain more realistic results.
For that we apply a non-local gaussian regulator
to the momentum space OKE potential
with a cut-off of the order of the hard scale 
($\Lambda = 0.5-1.0\,{\rm GeV}$),
which we plug into
the Lippmann-Schwinger equation~\cite{Nieves:2012tt,HidalgoDuque:2012pq}.
This choice is not the easiest one -- it generates a non-local potential --
but it is more convenient for a prospective
three body $DDK$ calculation.
Choices such as a monopolar or a gaussian form factor depending
on $\vec{q}$ will lead to a local potential
that can be used in the Schr\"odinger equation.

In the first scenario -- $D_{s0}^*$/$D_{s1}^*$ as a compact meson --
a $D D_{s0}^*$ ($D^* D_{s1}^*$) bound state is very likely but
the uncertainties are large.
For $h = 0.7$ the binding energy is $E_B = -(4-13)\,{\rm MeV}$, with the spread
reflecting the cut-off range. 
This figure decreases to $E_B = -(1-5)\,{\rm MeV}$ if we choose $f_K$
instead of $f_{\pi}$ in the OKE potential.
The system binds for most choices of the parameters
except for $h = 0.5$ with $f_K$ (though there is still
a virtual state at $E_V = -0.7\,{\rm MeV}$).
The resilience against short-range physics
can be illustrated by changing the regulator to
\begin{eqnarray}
V(r; R_c) = V(r)\,\theta(r - R_c) \, , \label{eq:OKE-reg}
\end{eqnarray}
where $R_c$ is a cut-off radius.
With this regulator OKE binds for
$R_c \leq 0.8-1.3\,{\rm fm}$ ($0.9-1.4\,{\rm fm}$)
depending on whether we use $f_{\pi}$ or $f_{K}$.
This is larger than the typical range of short distance physics,
which as we will see below also happen to be suppressed.
If we consider the exchange of other light mesons, we notice that
SU(3) flavour symmetry and the OZI rule imply that the coupling of
the $D_{s0}^*$/$D_{s1}^*$ to the sigma and omega mesons vanishes.
The only non-suppressed light meson exchange is that of the $K^*$,
which generates a spin-spin interaction that vanishes for $D_{s0}^* D$
while it is repulsive (attractive) for $S=0$ ($S=2$) $D_{s1}^* D^*$.
That is, OKE dominates the low energy physics of this system.

The importance of OKE can also be understood by reinterpreting the previous
predictions as the leading order (LO) calculation in an effective field
theory (EFT) with the heavy mesons and the pseudo Nambu-Golstone bosons
as the low energy degrees of freedom.
Within this framework the longest range correction
to the OKE potential comes from two pion exchange (TPE)~\footnote{
  Two kaon exchange does not benefit from the enhanced range of OKE and
  we do not further consider it.},
in particular the football and triangle diagrams~\cite{Machleidt:2011zz}.
These diagrams enter at $Q^2$ naively, where the $Q$ notation denotes the
ratio of a light scale (e.g. the pion mass or the effective kaon mass)
over a hard scale (e.g. the rho mass).
Yet they involve the $D_{s0}^* \pi \to D_{s0}^* \pi$ and
$D_{s1}^* \pi \to D_{s1}^* \pi$ amplitudes that vanish at lowest order,
demoting TPE to order $Q^3$.
In addition for the football diagram the lowest order $D\pi \to D\pi$ amplitude
cancels with $D_{s0}^* \pi \to D_{s0}^* \pi$ owing to their isospin structure.
As a consequence the football diagram is at least $Q^4$.
This is to be compared with OKE, which we count as $Q^{-1}$ following Refs.~\cite{Barford:2002je,Birse:2005um,Valderrama:2016koj}. 
In short, OKE is well-protected from subleading corrections.

%
%

In the second scenario -- the $D_{s0}^*$/$D_{s1}^*$ are molecular --
the couplings are rather well constrained and the binding energy
is amenable to error estimations.
Concrete calculations indicate the existence of a $D D_{s0}^*$ ($D^* D_{s1}^*$)
bound state with
$E_B = -4_{-5}^{+3}\,{\rm MeV}$ ($E_B = -5_{-5}^{+3}\,{\rm MeV}$)
for $\Lambda = 0.5\,{\rm GeV}$ and
$E_B = -13_{-13}^{+8}\,{\rm MeV}$ ($E_B = -15_{-13}^{+9}\,{\rm MeV}$)
for $\Lambda = 1.0\,{\rm GeV}$,
with other regulators yielding similar numbers~\footnote{
  For instance, a monopolar form factor in each kaon vertex
  with $\Lambda = 0.8\,{\rm GeV}$ and $1.6\,{\rm GeV}$ yields
  $B = -4^{+3}_{-7}\,{\rm MeV}$ ($-6^{+4}_{-8}\,{\rm MeV}$) and
  $B = -16^{+11}_{-17}\,{\rm MeV}$ ($-20^{+13}_{-20}\,{\rm MeV}$),
  where the cut-off is chosen to be $\Lambda > m_{\rho}$
  as usual for form factors.
  For $\Lambda \to \infty$ we have (independently of regulator) 
  $E_B = - 40_{-50}^{+30}\,{\rm MeV}$ ($-50_{-50}^{+30}\,{\rm MeV}$).
}.
For a sharp cut-off $R_c$ OKE binds
for $R_c \leq 1.3^{+0.3}_{-0.3}\,{\rm fm}$ ($1.4^{+0.3}_{-0.3}\,{\rm fm}$),
which is a factor of two larger than the mean square radius
$\sqrt{\langle r^2 \rangle} \sim 0.7\,{\rm fm}$ of a $D K$/$D^*K$ bound state,
from which we deduce that binding is a solid prediction.

Yet the binding momenta of the $D_{s0}^* D$ /$D_{s1}^* D$ molecules
is about $100-200\,{\rm MeV}$, comparable to that of a $DK$/$D^*K$ molecule.
This points to corrections from the underlying $D D K$ structure.
The interactions in the $D D K$ system are of a short-range nature,
which makes a Faddeev calculation simple in this system
(details will be provided in a future publication).
If we fix the $D K$ contact-range interaction to reproduce
the $D_{s0}^*$ pole and assume that there is no $D D$ interaction,
for $\Lambda = 0.5-1.0\,{\rm GeV}$ we find a bound state
at $E_B = -(44-57)\,{\rm MeV}$ below the $D_{s0} K$ threshold.
From rho and omega exchange we expect the $DD$ potential to be
repulsive at short distances. This is taken into account
by saturating the $DD$ contact-range coupling
by the exchange of these two mesons as in~\cite{Epelbaum:2001fm}.
In this case we find $E_B = -(15-28)\,{\rm MeV}$.
Finally the inclusion of relativistic kaon kinematics
(in the formalism of Refs.~\cite{Garcilazo:1984rx,Mathelitsch:1986ez,Valderrama:2018knt})
and the correct energy dependence of the Weinberg-Tomozawa term
for the $D K$ interaction
give modest corrections to the numbers above,
about $\Delta E_B = +5\,{\rm MeV}$ for a non-interacting $D D$ pair and
$\Delta E_B = +1\,{\rm MeV}$ if we include $D D$ repulsion.


Other interesting aspect of the $D D_{s0}^*$ and $D^* D_{s1}^*$ molecules
is their decays, which are given by the decays of their components
plus interference and binding effects (analogous to those
in the $D^0\bar{D}^0 \gamma$ and $D^0\bar{D}^0 \pi^0$
decays of the $X(3872)$~\cite{Voloshin:2003nt,Guo:2014hqa,Guo:2014taa}).
While the width of the $D^*$ is of the order of
$100\,{\rm keV}$~\cite{Anastassov:2001cw,Ahmed:2001xc},
the widths of the $D$, $D_{s0}^*$ and $D_{s1}^*$ are not
that well known experimentally (except for upper bounds).
Theory suggests that they are narrow:
$D$ only decays weakly while the $D_{s0}^*$ and
$D_{s1}^*$ decays require isospin violation, where estimates of the width
of the former range from a
few ${\rm keV}$~\cite{Colangelo:2003vg,Guo:2006fu,Guo:2006rp}
to about a pair of hundred of ${\rm keV}$ at most~\cite{Guo:2008gp}.
Besides the binding energy of the $D^* D_{s1}^*$ molecule precludes
the possibility of the $D^* \to D \pi$ decay, as chances are that
this molecule is below the $D D_{s1}^* \pi$ threshold.
From this we can conclude that the width of these two molecular states
is really narrow, well below $1\,{\rm MeV}$, in agreement with the original
expectations  about $QQ \bar q \bar q$ states~\cite{Manohar:1992nd}.

Probably the most effective way to produce the $D D_{s0}$ and $D^* D_{s1}^*$
molecules in experiments involves heavy ion collisions, the reason being
their double charm content.
The production yields for the theoretical $T_{cc}$ tetraquarks
($cc \bar q \bar q$) and other exotic hadrons have been estimated
for electron-positron~\cite{Hyodo:2012pm} and
heavy ion collisions~\cite{Cho:2017dcy}, where the predicted yields
may be reachable by the LHCb in the future (double
charm baryon production has been recently
achieved~\cite{Aaij:2017ueg}).
Yet we note that the production of double charm molecules is probably different
from the estimates above, which refer to the more compact
$T_{cc}$ tetraquarks.

The previous ideas also apply to the bottom sector, where
the $B_{s0}(5730)$ and $B_{s1}(5776)$ bottom-strange mesons
have been theorized to have a significant molecular component
and a similar binding energy as the $D_{s0}^*$ and $D_{s1}^*$
mesons~\cite{Guo:2006fu,Guo:2006rp,Altenbuchinger:2013vwa}
(they also appear in lattice QCD calculations~\cite{Lang:2015hza}).
They are theoretical however and have not been experimentally discovered yet.
If we consider the $B B_{s0}$ and $B^* B_{s1}^*$ molecules, the OKE potential
is identical to the one for charm mesons but the spectrum will be
more bound owing to the heavier reduced mass.
For $\Lambda = 0.5\,{\rm GeV}$ / $1\,{\rm GeV}$
we find a $B B_{s0}$ bound state at
$E_B = -14^{+6}_{-7}\,{\rm MeV}$ / $-40^{+20}_{-20}\,{\rm MeV}$.
The $B^* B_{s1}^*$ predictions are almost identical because the reduced mass
is nearly the same.
Owing to their wave number, a sizeable three-body component is possible.
For $h = 0.8$ an excited shallow S-wave state appears.
For P-wave there is a bound state with $E_B = -4^{+4}_{-2}\,{\rm MeV}$ /
$-14^{+7}_{-10}\,{\rm MeV}$.
The previous uncertainties only take into account the error
in the coupling $h$. The biggest uncertainty will come from the actual
location of the $B_{s0}$ and $B_{s1}^*$ states: the closer they are
to the $B K$ and $B^* K$ threshold, the longer the range of
the OKE potential and the more probable additional
bound states will be.

To summarize, the $D D_{s0}^*$ and $D^* D_{s1}^*$ systems
interact via a long-ranged kaon exchange Yukawa potential.
This potential is a consequence of the different parities and masses of
the $D$($D^*$) and $D_{s0}^*$($D_{s1}^*$) heavy mesons. It also provides
an excellent opportunity to predict the existence of bound states:
owing to the non-singular character of the Yukawa potential,
predictions do not crucially depend on arbitrary short-range
physics, though there is still a moderate dependence
on the cut-off.
We find that there must be bound states with a binding energy of
$5-15\,{\rm MeV}$
where the exact number should be fairly independent on whether
we have a $0^{-}$ $D D_{s0}^*$ or a $0^{-}$/$2^{-}$ $D^* D_{s1}^*$ molecule.
These predictions are robust against short-range dynamics,
partly because the latter are suppressed phenomenologically.
If the bound states become too deep their description probably
requires the inclusion of a three body component
($D D K$ and $D^* D^* K$ respectively).
This does not affect the prediction of bound states, only their location.
We expect the existence of similar bound states in the bottom sector, i.e.
$B B_{s0}$ and $B^* B_{s1}^*$.
They will be more bound and might have a richer spectrum than their charm
counterparts (there is probably a P-wave state and an excited
S-wave one), but we remind that the $B_{s0}$ and $B_{s1}^*$ heavy mesons
have not been observed yet in experiments.
The mechanism behind these molecules and the dual three body description
probably extends to other hadron systems, for instance $\Lambda(1405) N$,
$\Xi(1690) \Sigma$ and $\Lambda(1520) \Sigma^*$
to name a few prominent examples.

\section*{Acknowledgments}

M.P.V thanks the Institut de Physique Nucl\'eaire d'Orsay, where part
of this work was done, for its hospitality,
This work is partly supported by the National Natural Science Foundation
of China under Grants No. 11375024,  No.11522539, No.11735003,
the Fundamental Research Funds for the Central Universities,
the Thousand Talents Progam for Youth Professionals
and by the JSPS KAKENHI Grant Number JP16K17694.

%

\end{document}